\documentclass[preprint]{aastex}

\usepackage{amssymb}
\usepackage{amsmath}
\usepackage{url}
\usepackage{epsfig}

\shorttitle{Electron Conduction at high $\beta$}
\shortauthors{Roberg-Clark et al.}

\usepackage{geometry}                
\usepackage{graphicx}
\usepackage{amssymb}
\usepackage{epstopdf}

\begin{document}

\title{Suppression of electron thermal conduction in the high $\beta$
  intracluster medium of galaxy clusters}

\author{G.~T.~Roberg-Clark\altaffilmark{1}, J.~F.~Drake\altaffilmark{2,5,6,7},
  C.~S.~Reynolds\altaffilmark{3,7}, M.~Swisdak\altaffilmark{4,6,7}}

\altaffiltext{1}{Department of Physics, University of Maryland,
  College Park, MD 20742, USA; grc@umd.edu}

\altaffiltext{2}{Department of Physics, University of Maryland,
  College Park, MD 20742, USA; drake@umd.edu}

\altaffiltext{3}{Department of Astronomy, University of Maryland,
  College Park, MD 20742, USA; chris@astro.umd.edu}

\altaffiltext{4}{Department of Physics, University of Maryland,
  College Park, MD 20742, USA; swisdak@umd.edu}

\altaffiltext{5}{Institute for Physical Science and Technology,
  University of Maryland, College Park, MD 20742, USA}

\altaffiltext{6}{Institute for Research in Electronics and Applied
  Physics, University of Maryland, College Park, MD 20742, USA}

\altaffiltext{7}{Joint Space-Science Institute (JSI), College Park, MD
  20742, USA}

\begin{abstract}
Understanding the thermodynamic state of the hot intracluster medium
(ICM) in a galaxy cluster requires a knowledge of the plasma transport
processes, especially thermal conduction. The basic physics of thermal
conduction in plasmas with ICM-like conditions has yet to be
elucidated, however.  We use particle-in-cell simulations and analytic
models to explore the dynamics of an ICM-like plasma (with small
gyroradius, large mean-free-path, and strongly sub-dominant magnetic
pressure) induced by the diffusive heat flux associated with thermal
conduction.  Linear theory reveals that whistler waves are driven
unstable electron heat flux, even when the heat flux is weak.  The
resonant interaction of electrons with these waves then plays a
critical role in scattering electrons and suppressing the heat
flux. In a 1D model where only whistler modes that are parallel to the magnetic field are captured, the only resonant
electrons are moving in the opposite direction to the heat flux and
the electron heat flux suppression is small. In 2D or more, oblique
whistler modes also resonate with electrons moving in the direction of
the heat flux. The overlap of resonances leads to effective
symmetrization of the electron distribution function and a strong
suppression of heat flux.  The results suggest that thermal conduction
in the ICM might be strongly suppressed, possibly to negligible
levels.
\end{abstract}


\section{Introduction}

Over 80\% of the baryonic matter in a galaxy cluster resides in an
atmosphere of hot plasma, the intracluster medium (ICM), which is in a
state of approximate hydrostatic equilibrium within the gravitational
potential of the cluster's dark matter halo. In many
clusters, X-ray measurements of the electron number density ($n_e\sim
10^{-3}-10^{-1}\,{\rm cm}^{-3}$) and temperature ($T\sim
10^7-10^8\,{\rm K}$) reveal ICM cores that have short cooling times
($t_{\rm cool}<10^9\,{\rm yr}$) and depressed temperatures \citep{fabian94a}.  If
unchecked, the radiative losses in these cool-core clusters would lead
to significant accumulations of cold gas within the central galaxy,
resulting in star formation rates of $100-1000\,{\rm M}_\odot\,{\rm
  yr}^{-1}$, and central galaxies with stellar masses of
$10^{13}\,{\rm M}_\odot$ or more \citep{croton06a}. Observed star formation rates and
total stellar masses in these systems are an order of magnitude
smaller, demonstrating that the radiative losses of the ICM must be
largely offset. The current paradigm is that energy injection by a
central jetted active galactic nucleus (AGN) is thermalized in the
ICM \citep{churazov00a,churazov02a,reynolds02a}. Thermal conduction within the ICM is very likely to play a
central role in these astrophysical processes by dissipating weak
shocks and sound waves driven by the AGN, and strongly modifying local
thermal instabilities \citep[and references therein]{binney81a,fabian05a,yang16a}.  The direct transport of heat from the outer
(hotter) regions of the ICM may also be a source of heat for the ICM
cool-core.

Thermal conduction in the ICM plasma remains poorly understood.
At these densities and temperatures, the electron mean
free path is $\lambda\sim 0.1-1\,{\rm kpc}$. With measured magnetic
fields of $B\sim 1-10\mu{\rm G}$, the electron gyro-radius $\rho_e\sim
10^8\,{\rm cm}$ is many orders of magnitude smaller so
transport is highly anisotropic.  Most current treatments of the ICM
adopt a fluid description, taking the
thermal conductivity to have the canonical Spitzer value
\citep{spitzer56a} along the local magnetic field and
complete suppression in the orthogonal direction. However, Spitzer
conductivity is not likely to be valid in the low collisionality ICM
plasma where collisional mean-free-paths and temperature scale lengths
can be comparable.  Furthermore, the fact that the ratio of
thermal-to-magnetic pressure is large, $\beta\equiv 8\pi n T/ B^2 \sim
100$ suggests that the ICM is susceptible to instabilities
driven by pressure anisotropies and heat fluxes that are expected
to impede thermal conduction \citep{gary00a,li12a,kunz14a,rincon15a}.

In this paper we explore how self-generated turbulence impacts the
thermal conductivity of a high-$\beta$ ICM plasma. Unlike in earlier
models in which the pressure anisotropy in the high-$\beta$ medium is
a source of turbulence that impacts thermal conduction
\citep{riquelme16a,komarov2016a}, we focus directly on the electron
heat flux as a source of free energy. We show that whistler waves
driven by the heat flux are generically unstable in the
ICM. Particle-in-cell (PIC) simulations of the turbulence
reveal strong heat-flux suppression. A comparison of the results of 1D
and 2D simulations with an analytic model reveals the importance of
the resonant interaction between the electrons and waves and
associated particle trapping in facilitating strong scattering.

\section{1D Instability Model}

We solve the linearized Vlasov-Maxwell equations to obtain a
dispersion relation for whistler-like modes propagating along the
local magnetic field $\mathbf{B} = B_0 \hat{\mathbf{x}}$. The temperature $T$ is taken to be uniform but we include
a heat flux as a source of free energy. We neglect ion contributions, which scale like $\sqrt{m_e/m_i}$ and are small in simulations (not shown). Making the standard whistler assumptions, we obtain the dispersion relation for the frequency $\omega$ of modes with wave vector $k$ along $\mathbf{B}$,
\citep{krall86c}:
\begin{equation}\label{KTdisp}
\frac{k^2c^2}{\omega^2} - \frac{\omega_{pe}^2}{\omega n_{0}} \int
d^3\mathbf{v}\ \frac{ v_{\perp}}{2}
\frac{\left[\left(1-\frac{kv_x}{\omega} \right)
    \frac{\partial f_{0}}{\partial v_{\perp}} + \frac{k
      v_{\perp}}{\omega} \frac{\partial f_{0}}{\partial v_x}
    \right]}{\omega - kv_x - \Omega_e} = 0
\end{equation}
where $\omega_{pe} = (4\pi n_0 e^2/m_e)^{1/2}$ is the plasma
frequency, $\Omega_e = eB_0/m_{e}c$ is the cyclotron frequency,
$f_0(\mathbf{v})$ is the initial electron phase space distribution,
$v_{Te} =(2T_e/m_e)^{1/2}$ is the thermal speed, $\rho_e =
v_{Te}/\Omega_e$ is the Larmor radius, $d_e = c/\omega_{pe}$ the skin
depth and $\beta_e = 8\pi n_{0} T_e/B^2 $.

In the standard whistler ordering with $\omega \sim (kd_e)^2 \Omega_e
\sim \Omega_e$ and $kd_e \sim 1$, waves in high beta plasmas resonate
with bulk electrons, $v_x \sim \Omega_e / k \sim \Omega_e d_e \sim
v_{Te} / \sqrt{\beta_e} \ll 1$, and are therefore heavily damped. Thus
whistlers with conventional ordering do not exist in high beta
plasmas. To obtain wave growth we consider longer wavelength modes
with $\omega \sim (kd_e)^2 \Omega_e \ll \Omega_e$. Resonant particles
have $v_x \sim \Omega_e / k \sim v_{Te} / (k\rho_e)$. Requiring $v_x
\gtrsim v_{Te}$ yields $k\rho_e \lesssim 1$ with $\omega \sim \Omega_e /
\beta_e \ll \Omega_e$.

\begin{figure}
\plotone{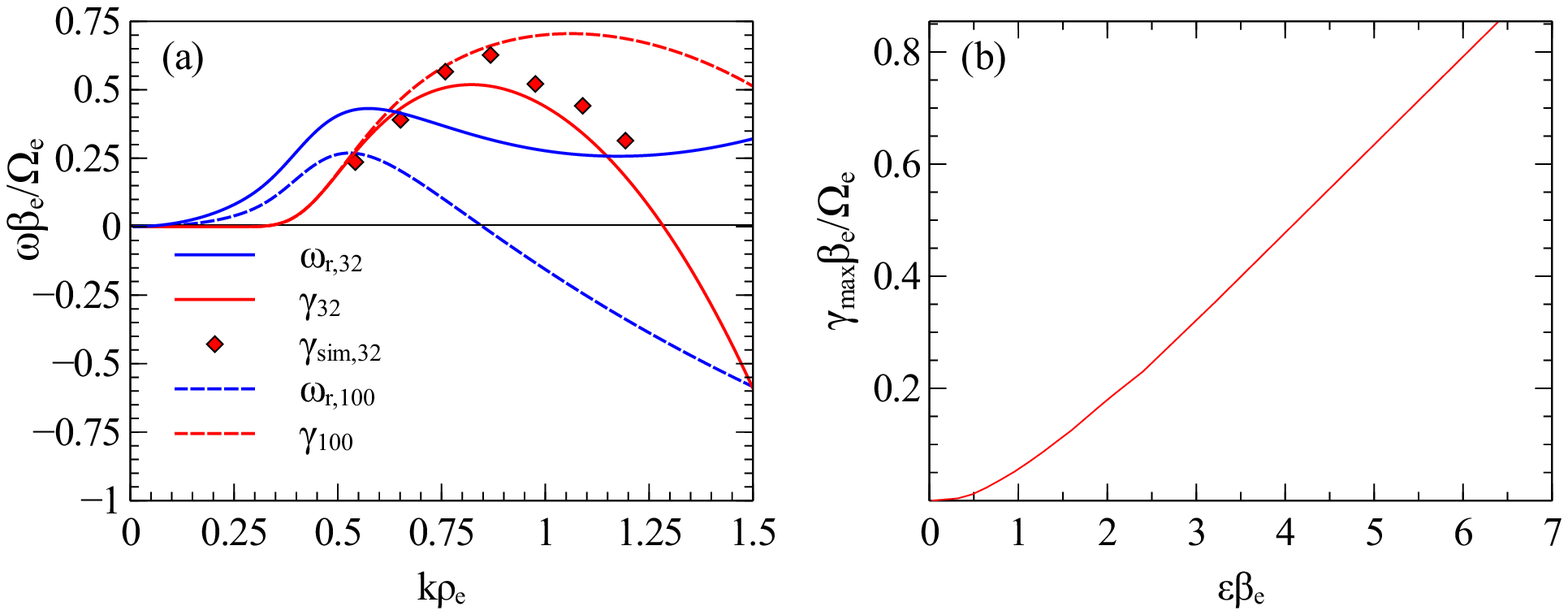}
\caption{Analytic dispersion relation of the heat-flux-driven
  whistler-like wave in a plasma with $\beta_{e} = 32 \ \text{(solid lines)}, 100 \ \text{(dashed lines)}$ and $\epsilon =
  0.133$. (a) Real frequency (blue) and growth rate (red) of the
  instability as calculated from Eq.~(\ref{1ddisp}). The red diamonds
  show growth rates at discrete values of $k$ taken from a 1D
  simulation with the same $\epsilon$ and $\beta=32$. (b) Growth rate of
  the maximally growing mode for a range of $\epsilon\beta$. }
\label{fig:1}
\end{figure}

To model heat flux instability we use a
distribution function from \cite{levinson92a}
\begin{equation}\label{LE}
f_0(\mathbf{v}) = f_{m}\left[1 + \epsilon \left(\frac{v^2}{v_{Te}^2} -
  \frac{5}{2} \right) \frac{v_x}{v_{Te}}\right]
\end{equation}
where $f_{m} = n_{0e}/(\sqrt{\pi}v_{Te})^3 \: \text{exp}
\left[-v^2/v_{Te}^2\right]$ and the term proportional to $\epsilon =
v_{Te}/ (\nu_{ei} L_T ) \ll 1$ yields a heat flux (see also
\cite{ramani78a}). Equation \ref{LE} was obtained by balancing a large-scale
temperature gradient (along $\mathbf{B_0}$) $\partial T / \partial x
\equiv T / L_T$ with a Krook collision operator. $f_0$ has no
net drift ($\langle \mathbf{v} \rangle = 0$) and the plasma pressure
is isotropic. The sole driver for instability is the heat flux,
$q_{x0} = mn_0\langle v_xv^2 \rangle /2 = (5/8)\epsilon m
n_0v_{Te}^3$. Using this distribution function and taking $\omega \ll
\Omega_e$ in Eq. \ref{KTdisp} the frequency becomes
\begin{equation}\label{1ddisp}
\omega = \frac{\Omega_e}{\beta_e}\left(k^2\rho^2_e + \epsilon \beta_e
k \rho_e h_2 \right) \frac{1}{h_1},
\end{equation}
where 
\begin{equation}
h_1(k\rho_e, \epsilon) = \frac{\Omega_e}{kn_0} \int d^3\mathbf{v}
\frac{f_0}{v_x + \Omega_e/k} \nonumber
\end{equation}
and 
\begin{equation}
h_2(k\rho_e)= \frac{1}{k\rho_e} \frac{1}{\epsilon} \int d^3 \mathbf{v}
\frac{v_xf_0 + (v_{\perp}^2/2)\partial f_0 / \partial
  v_x}{v_x + \Omega_e / k}. \nonumber
\end{equation}
$h_2$ is only non-zero because of the heat flux and the integral over
$v_x$ must go under the singularity at $v_x =
-\Omega_e/k\equiv v_r$, the parallel resonant velocity. The real
frequency $\omega_r$ and growth rate $\gamma$ versus $k$ for $\beta_e
= 32,100$ and $\epsilon = 0.133$ are presented in Fig. \ref{fig:1}(a). The
growth rate $\gamma$ is peaked around $k\rho_e \sim 1$ and goes to
zero for small and large $k\rho_e$. The waves have whistler-like
dispersion for small $k$ but the frequency rolls over at $k\rho_e
\simeq 0.6$ and has a characteristic phase speed $ v_{ph} = \omega /
k \sim \rho_e \Omega_e / \beta_e = v_{Te} / \beta_e \ll v_{Te}$. The
resonant interaction is with particles with $v_x\sim -v_r\sim
-v_{te}$.  In Fig. \ref{fig:1}b the maximum growth rate (with respect
to $k$) is plotted against $\beta\epsilon$. Instability exists for any
non-zero value of $\epsilon$ so there is no threshold for the instability. This is highly relevant to the ICM, for which
$\epsilon \ll 1$ \citep{levinson92a}. We now
present numerical simulations to verify
(\ref{1ddisp}) and to probe the impact on the heat
flux.

\section{Numerical Methods}

We simulate the instability using the particle-in-cell (PIC) code
$\tt{p3d}$ \citep{zeiler02a}. Particle trajectories are calculated
using the relativistic Newton-Lorentz equations and the
electromagnetic fields are advanced using Maxwell's equations. We
present the results of quasi-1D, collisionless simulations (3600
particles per cell) with dimensions $L_x \times L_y = 28.96 \rho_e
\times 1.80 \rho_e $ (a finite $L_y$ increases the number of
particles and reduces particle noise) and a 2D simulation with $L_x
\times L_y = 28.96 \rho_e \times 28.96 \rho_e$ (800 particles per
cell). Periodic boundary conditions are used in both $x$ and $y$ with
$\beta_{e0}=32$. With these values of $L_x$ and $L_y$ many unstable modes
of scale $\rho_e$ can fit in the box. Ions form a stationary,
charge-neutralizing background.

There is no ambient temperature gradient but we initialize
electrons with the distribution given in Eq.~\ref{LE}. Since $f_0$ is not strictly positive we adjust our initial distribution
to ensure that $f_{0}\geq 0$ and that it has no net drift or pressure
anisotropy. Since $q_{x0}$ is also affected, we calculate an effective
initial $\epsilon$ for comparison with the stability theory. Our 1D
simulations are run to $t=24.4 \ \beta_{e0}/\Omega_{e0}$ and a single
2D simulation is run to $t = 30.6 \ \beta_{e0}/\Omega_{e0}$.

\section{1D Simulation Results}

\begin{figure}
\plotone{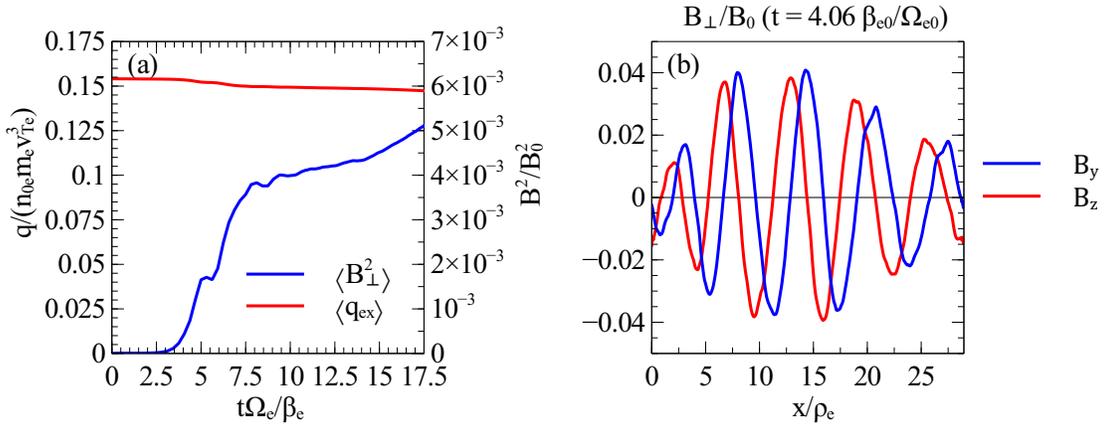}
\caption{1D PIC Simulation Results. (a) Average heat flux and mean
  squared value of the perturbed magnetic field. (b) Whistler-like
  phase relation between $B_{y}$ and $B_{z}$.}
\label{fig:2}
\end{figure}

The 1D simulation, which has an effective $\epsilon = 0.246$, reveals
that the heat flux drives waves unstable as predicted by
Eq.~\ref{1ddisp}. Magnetic fluctuations perpendicular to
$\mathbf{B_0}$ grow in time (Fig. \ref{fig:2}a) and saturate with
$\tilde{B}_{sat} \simeq 0.1 \: B_0$ . The waves are right-hand
circularly polarized with $B_y$ and $B_z$ $90^{\circ}$ out of phase
(Fig.~\ref{fig:2}b). The linear growth rates from the simulation are
in good agreement with those obtained from the linear theory
(Fig. \ref{fig:1}a).

\begin{figure}
\plotone{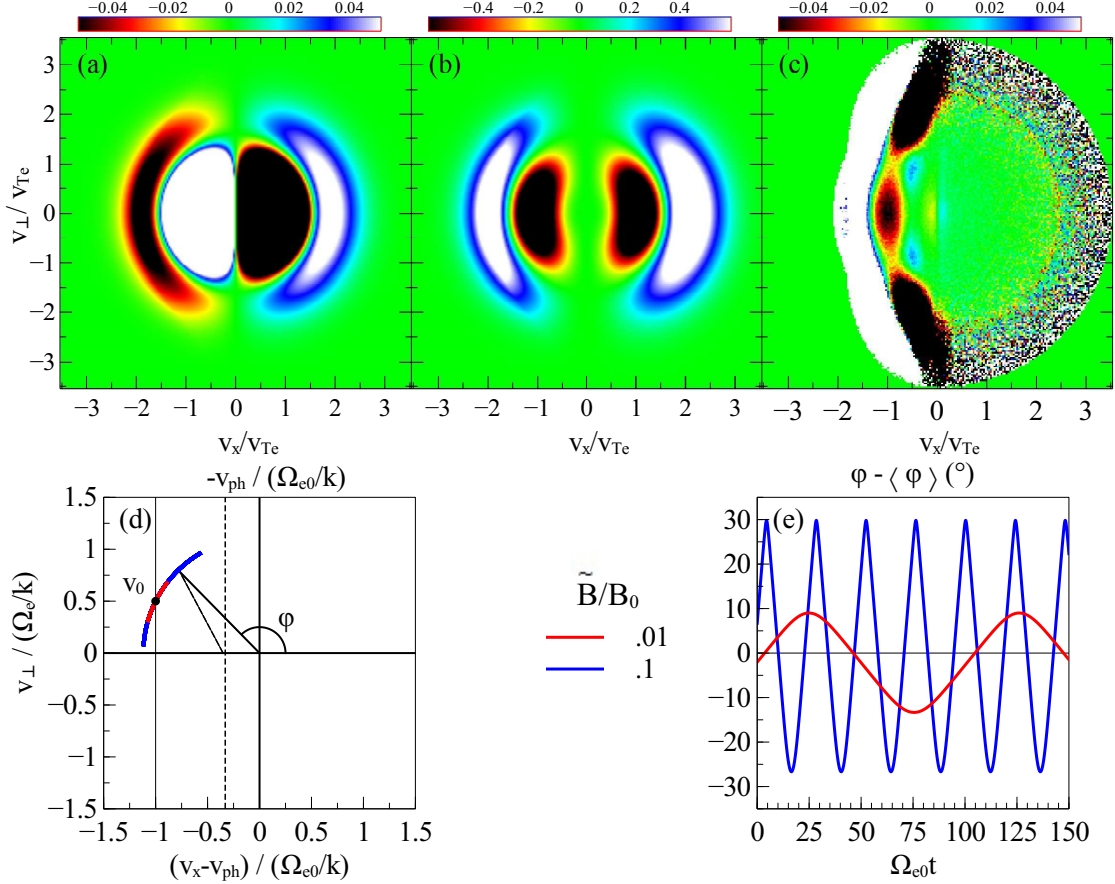}
\caption{Phase space plots and particle trapping in 1D. (a) $(f_0-f_M)/\text{max}(f_0-f_M)$ from the PIC simulation. (b) The local heat flux,
  $(f_{0}-f_{M}) v^{2} v_x /\text{max}[(f_{0}-f_M) v^{2} v_x]$. (c) $\delta
  f/f_0=(f-f_0)/f_0$ from the PIC simulation
  at late time, $t=9.34 \ \beta_{e0}/\Omega_{e0}$. (d) Trajectories of trapped test particles in the
  wave frame of the 1D whistler for small (red) and large (blue)
  $\tilde{B}$. The initial parallel velocity is the resonant velocity
  $-\Omega/k$.  The lab frame is marked with the dashed lines. In
  (e) temporal evolution of the angle $\phi=\arctan (v_\perp /v_x)$
  for the test particles in (d). \label{fig:3}}
  
\end{figure}

The instability's nonlinear evolution coincides with a surprisingly weak reduction of the total heat flux
(Fig.~\ref{fig:2}a). The reason for this behavior is linked to the
mechanism by which whistlers gain energy from particles and are
then scattered to reduce the heat flux. In the frame moving with the
wave, particles move along concentric circles of
constant energy (see Fig.~\ref{fig:3}d).  Resonant electrons moving
from high $v_\perp$ to low $v_\perp$ along the constant energy contour
in the wave frame lose energy in the simulation frame. If more
particles move in this direction than toward higher $v_\perp$, the
waves grow. The portion of the distribution function proportional to
$\epsilon$ in Eq.~\ref{LE} and the associated heat flux are shown in
Figs.~\ref{fig:3}a,b. Resonant particles that drive instability
have $v_\perp >1.5$. This picture is confirmed in Fig.~\ref{fig:3}c,
which shows the change in the electron distribution function as a result of the instability. Note the depletion of
electrons with high $v_\perp$ and negative $v_x$. Saturation occurs
when this relatively small region is depleted of excess
particles. Figure \ref{fig:3}b indicates that the bulk heat flux in
phase space is carried by $v_x > 0$ particles with high energy, where
the distribution function in Fig.~\ref{fig:3}c is essentially
unchanged. Since positive velocity electrons cannot
resonate with the 1D whistler instability, significant heat flux
suppression cannot occur.

 \section{1D Trapping Model}
Modest heat flux reduction in 1D is linked to constraints on how
electrons are scattered in a collisionless system. In 1D it is only
electrons with velocity $v_x=-\Omega /k$ that resonantly drive the
instability (Fig. \ref{fig:3}c) and we will show that it is only
particles close to this resonance that scatter. The substantial number
of particles with $v_x>0$, which carry the bulk of the heat flux
(Fig.~\ref{fig:3}b,c), do not participate in heat flux
suppression. The importance of resonant interactions and particle
trapping in the scattering of particles by waves in magnetized plasma
has been discussed by Karimabadi {\it et al}, 1992 based on a formal
Hamiltonian theory \citep{karimabadi92a}.

We demonstrate this here by considering electrons in a whistler
propagating in the positive $x$ direction, $\tilde{{\bf
    B}}=\tilde{B}(\hat{\mathbf{y}}\sin(kx-\omega
t)+\hat{\mathbf{z}}\cos(kx-\omega t))$. In the frame moving with the
whistler, the electric field is zero and the energy,
$v_x^2+v_y^2+v_z^2=v_0^2$, is conserved \citep{karimabadi92a}. The
equation of motion in the wave frame is
\begin{equation}
\frac{d{\bf v}}{dt}=-\Omega_e {\bf
  v}\times\hat{\mathbf{x}}-\Omega_e\frac{{\bf v}\times\tilde{{\bf
      B}}}{B_0}.
\label{eqn:vdot}
\end{equation}
The fast time variation of the cyclotron motion can be eliminated by
defining the new variables $v_\pm =(v_y\pm iv_z)e^{\mp i\Omega_e t}$,
\begin{equation}
\frac{dv_\pm}{dt}=v_x\tilde{\Omega}_e e^{\mp i(kx-\omega t)}
\label{eqn:vpm1}
\end{equation}
where $\tilde{\Omega}_e=e\tilde{B}/m_ec$. Shifting to a moving frame
with velocity $-\Omega_e /k$, we define $\bar{x}=x+\Omega_e t/k$ with
$\bar{v}_x=v_x+\Omega_e /k$ so that Eq.~\ref{eqn:vpm1} becomes
\begin{equation}
\frac{dv_\pm}{dt}=\left(\bar{v}_x-\frac{\Omega_e}{k}\right)\tilde{\Omega}_{e}e^{\mp
  ik\bar{x}}
\label{eqn:vpm2}
\end{equation}
with energy conservation now given by $v_+v_-+(\bar{v}_x-\Omega_e
/k)^2=v_0^2$. The time variation of the particles in this frame is completely
controlled by $\tilde{\Omega}_e$. The phase variation of the whistler,
however, limits the excursion of $v_\pm$. As $v_\pm$ increases or
decreases, $\bar{v}_x$ changes due to energy conservation and the wave
phase $k\bar{x}$ changes even if $\bar{v}_x$ were initially zero. Consequently, the change in $v_\pm$ eventually reverses and the
electrons are trapped (Fig. \ref{fig:3}d). To show this,
 we take the time derivative of the energy relation and use
Eq. \ref{eqn:vpm2} to obtain an equation for
$\dot{\bar{v}}_x=\ddot{\bar{x}}$,
\begin{equation}
\ddot{\bar{x}}+\tilde{\Omega}v_\perp\cos(k\bar{x}+\phi)=0,
\label{eqn:xddot}
\end{equation}
where we have written $v_\pm=v_\perp e^{\pm i\phi}$. Because trapping
limits the excursion of $v_\pm$ we can approximate $v_\perp$ and
$\phi$ by their initial values $v_{\perp 0}$ and $\phi_0$. The
equation for the phase angle $\theta =k\bar{x}+\phi_0+\pi /2$ is
\begin{equation}
\ddot{\theta}+\omega_b^2\sin\theta=0,
\label{eqn:theta}
\end{equation}
where $\omega_b=\sqrt{kv_{\perp 0}\tilde{\Omega}_e}$ is the bounce
frequency associated with deeply trapped particles. Integrating once
yields
\begin{equation}
\frac{1}{2}\dot{\theta}^2-\omega_b^2(1-\cos\theta)=\frac{1}{2}\dot{\theta}_0^2
\label{eqn:energy}
\end{equation}
where $\dot{\theta}_0$ is the value of $\dot{\theta}$ at $\theta
=0$. The maximum excursion of $\dot{\theta}$ corresponds to the
separatrix in the phase space of $\theta-\dot{\theta}$ which is
defined by $\dot{\theta}_0=0$. Thus, $\Delta\dot{\theta}=2\omega_b$ is
the trapping width. This corresponds to excursions
\begin{equation}
\frac{\Delta v_x}{v_r}=2\sqrt{\frac{v_{\perp
      0}}{v_r}\frac{\tilde{B}}{B_0}}\ll 1
\label{eqn:trapx}
\end{equation}
in $v_x$ and
\begin{equation}
\frac{\Delta v_\perp}{v_r}=2\sqrt{\frac{v_r}{v_{\perp
      0}}\frac{\tilde{B}}{B_0}}\ll 1
\label{eqn:trapperp}
\end{equation} 
in $v_\perp$, where $v_r=\Omega_e /k$. The same excursion was
calculated for ions moving in circularly polarized Alfv\'en waves
\citep{mace12a,dalena12a}. These bounds define the region in velocity
space where electrons are scattered. Electrons outside of these
ranges, which includes all of the positive velocity particles that
carry the bulk of the heat flux, are not scattered.

To confirm the predictions of the trapping theory, we initialize a particle in the frame of a whistler-like wave in 1D with a resonant parallel velocity $v_x=-\Omega_{e}/k$. The particle trajectory in azimuthal angle $\phi=\arctan
(v_\perp /v_x)$ in Fig.~\ref{fig:3}e reveals the trapped
bounce motion. The corresponding excursion is shown in
Fig.~\ref{fig:3}d.  As predicted, both the frequency and amplitude of
oscillations depend on $\tilde{B} / B_{0}$. Particles outside of
resonance (not shown) exhibit only small amplitude oscillations.

\section{2D Simulations and Analytic Theory}

\begin{figure}
\plotone{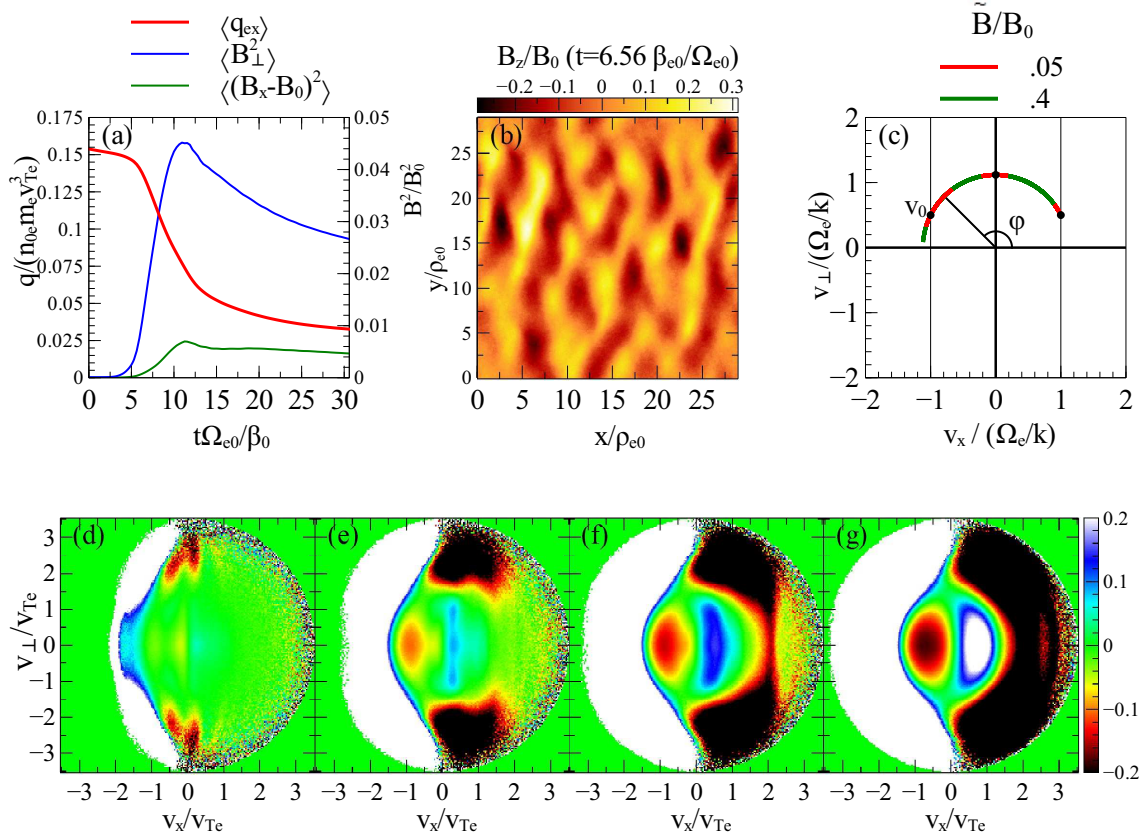}
\caption{2D simulation results and trapping theory. (a) Averages of the heat
  flux and mean-squared values of the perturbed perpendicular and parallel magnetic fields
  versus time. (b) The 2-dimensional structure of the magnetic
  field $B_z$. (c) In the wave frame of a 2D whistler, the orbits
  of trapped test particles for two values of $\tilde{B}$. The $v_x-v_\perp$ phase space of $\delta f/f_0 = (f-f_0)/f_0$ at
   $t\Omega_{e0}/\beta_{e0} = 5.31$ (d), $8.12$ (e), $10.94$ (f), and $30.6$ (g).}
  
\label{fig:4}
\end{figure}

In contrast with the 1D simulations, the suppression of heat flux in
2D is substantial: for an initial $\epsilon = 0.246$ the heat flux decreases to
$\approx 25\%$ of its starting value
(Fig. \ref{fig:4}a). Perturbations grow at a rate similar to those in
the 1D case, have $k\rho_e \sim 1$, and propagate ($\omega / k
\simeq v_{Te} / \beta_e$) along and
perpendicular to the magnetic field (Fig. \ref{fig:4}b) with a
characteristic $k_{\perp} = k_y \lesssim k_x$. The saturation time in
2D is only slightly longer than that in 1D. At $t\Omega_e /
\beta_e \simeq 12.5$ the amplitude of magnetic perturbations reaches
$0.4 B_0$. A time sequence of $\delta f / f_0$ (Fig. \ref{fig:4}d-g)
shows the development of resonances that were not present
in the 1D system. Measured resonant velocities $v_{x,\text{res}}$ are
a consequence of $k_\perp \neq 0$ and are given by the condition
$\omega - k_xv_x - n \Omega = 0 \nonumber$ where $n$ is any integer
\citep{krall86c,karimabadi92a}. In particular, we observe that the
$n=1$ $(v_x < 0)$, $n=0$ $(v_x \simeq 0)$ and $n=-1$ $(v_x > 0)$
resonances all play crucial roles. At the point of
saturation $f$ is significantly more isotropic in phase space than
$f_0$ (not shown). Furthermore, the region of concentrated heat flux
in Figure \ref{fig:3}b ($v_x \gtrsim v_{Te}$) has been drained of
excess particles. Whereas in 1D the trapping mechanism was unable to
significantly reduce the heat flux, in 2D the availability of the
$n=0$ and $n=-1$ resonances and resonant overlap
allow trapping to drive strong pitch-angle scattering over a broad
range of velocities. This scattering isotropizes the distribution
function by connecting the $v_x < 0$ and $v_x > 0$ regions of phase
space. 

Trapping equations in the 2D case were derived in \cite{karimabadi90a}
and the results are similar in form to
(\ref{eqn:theta}) and (\ref{eqn:energy}). The results for the $n=0$
(Landau) and $n=\pm 1$ (Cyclotron) resonances are nearly identical to
the 1D case and have the form of Equation \ref{eqn:theta}, where
$\omega_b \simeq \sqrt{kv_{\perp0}\tilde{\Omega}_e}$.  To demonstrate
the importance of these resonances, we evaluate test
particle orbits in the reference frame of a 2D off-angle whistler
wave. In this frame the wave takes the form
\begin{equation}
\mathbf{B}= B_0\hat{\mathbf{x}}+\tilde{B}\left[ \left(-
  \frac{k_y}{k_x}\hat{\mathbf{x}}+\hat{\mathbf{y}}\right)\sin(k_x x +
  k_y y)+\frac{k}{k_x}\hat{\mathbf{z}}\cos(k_x x + k_y y)\right]
\nonumber
\end{equation} 
where $\tilde{B}_x\neq 0$ and the wave is elliptically, rather than
circularly, polarized. In evaluating the particle orbits we take $k_y
= k_x$ and consider two different values for $\tilde{B} / B_0$: $0.05$
and $0.4$. Particles are initialized with parallel velocities at the
$n = 0,\pm 1$ resonances. In the case of $\tilde{B} / B_0 = 0.05$ we
find that trapping widths are small, as predicted by the nonlinear
theory (Figure \ref{fig:4}c). However, when $\tilde{B} / B_0 = 0.4$, a
particle starting at $v_x = \Omega_e / k $ experiences strong
trapping and is scattered into the domains of the other two resonances,
reversing its original parallel velocity. This is consistent with the results of \cite{karimabadi92a}, in which it was found that resonant overlap occurs when $\delta{B}/B_{0} \simeq 0.3$. It is not a coincidence that saturation takes place once perturbed amplitudes of this size are reached, for it is by this
mechanism, in which particles are ``handed off'' between different
resonances, that the particles driving instability are scattered in
phase space and wave growth ceases.

\section{Conclusions}

We have shown using both PIC simulations and linear theory that
low-frequency ($\omega\sim \Omega_e/\beta_e$) whistler-like modes in a
high-$\beta$ collisionless plasma are driven unstable by thermal heat
flux, even when the pressure is isotropic.  The non-linear
suppression of the heat flux is negligible in 1D, but becomes
substantial in 2D owing to overlapping Landau and
cyclotron resonances that lead to effective scattering of
electrons. This strong suppression of thermal heat
fluxes may be important for understanding the thermodynamics of
the ICM in galaxy clusters.

In order to quantify the astrophysical importance of these effects, we
calculate an effective conductivity that can be implemented into
global, fluid models of the ICM atmosphere. The electron distribution
function was earlier calculated by balancing the background
temperature gradient with a Krook collision operator, producing
the distribution function given in Eq.~(\ref{LE}), where the term
proportional to $\epsilon=v_{Te}/\nu_{ei}L_T$ describes the heat flux
\citep{levinson92a}. Although we have not carried out a scaling study
of the rate of wave-driven electron scattering $\nu_w$ with
parameters, a reasonable hypothesis is that $\nu_w$ is given by the peak linear
growth rate $\nu_w=\gamma=\epsilon\Omega_e$. Repeating the heat flux
calculation by replacing $\nu_{ei}$ with $\nu_w$ yields
$\epsilon=\sqrt{\rho_e/L_T}$ and a heat flux $q_{\parallel}$ given by
\begin{equation}
  q_{\parallel} \propto v_{te}nT_e\sqrt{\frac{\rho_e}{L_T}}.
  \label{eqn:heatflux}
\end{equation}
This is reduced from the collisionless, free-streaming value by the
factor $\sqrt{\rho_e/L_T}$, which can be as small as $10^{-6}$ for the
ICM.

A caveat is that our numerical models are run with large heat fluxes
($\epsilon\approx 0.25$) whereas typical heat fluxes in the ICM can be
much smaller. While linear theory shows that the whistler instability
exists for any non-zero heat flux, it is possible that there exists a
threshold heat flux below which whistler-mediated scattering of
electrons is ineffective. On the other hand, our present simulations
are initial value problems that relax to an equilibrium system in a
periodic box, while a more realistic configuration would continually
drive a heat flux down a temperature gradient, explicitly linking heat
flux and temperature gradient. The unlimited supply of free energy
into the system would likely drive the large-amplitude perturbations
that satisfy the resonance overlap condition, $\delta{B}/B_0 \gtrsim
0.3$, regardless of how small the driving heat flux is. This will
likely lead to a saturated state in which injection and disruption of
heat flux balance and so differ significantly from that of the initial
value problem in which heat flux relaxes to a low level. Whistler
turbulence from the heat flux instability might also be damped or
driven unstable by electron pressure anisotropies in the ICM and may
couple to ion-scale anisotropy-driven modes. These issues will be
explored in future work.

\begin{acknowledgments}

The authors wish to acknowledge NSF grant AST1333514 and NASA/SAO
Chandra Theory grant TM617008X, and M.W. Kunz for helpful discussion.

\end{acknowledgments}


\end{document}